\documentclass[ a4paper, twocolumn, pra, superscriptaddress, amsmath, amssymb, floatfix]{revtex4-1}
\usepackage{graphicx}        
\usepackage{amssymb}   
\usepackage{amsmath}
\usepackage{color,soul}
\usepackage{dcolumn}
\usepackage{bm}
\usepackage{mathtools}
\usepackage[table]{xcolor}
\usepackage{epstopdf}
\usepackage{float}

\newcommand{\ket}[1]{{\vert}#1{\rangle}}
\newcommand{\bra}[1]{{\langle}#1{\vert}}
\newcommand{\e}{\mathrm{e}}

\begin{document}

\title{Thermodynamic fingerprints of non-Markovianity in a system of coupled superconducting qubits}
\author{Sina Hamedani Raja}
\affiliation{QTF Centre of Excellence, Turku Centre for Quantum Physics, Department of Physics and
Astronomy, University of Turku, FI-20014 Turun yliopisto, Finland}

\author{Massimo Borrelli}
\affiliation{QTF Centre of Excellence, Turku Centre for Quantum Physics, Department of Physics and
Astronomy, University of Turku, FI-20014 Turun yliopisto, Finland}

\author{Rebecca Schmidt}
\affiliation{QTF Centre of Excellence, Turku Centre for Quantum Physics, Department of Physics and
Astronomy, University of Turku, FI-20014 Turun yliopisto, Finland}
\affiliation{Center for Quantum Engineering, Department of Applied Physics,
Aalto University School of Science, P.O. Box 11000, FIN-00076 Aalto, Finland}

\author{Jukka P. Pekola}
\affiliation{7 QTF Centre of Excellence, Department of Applied Physics, Aalto University, FI-00076
Aalto, Finland}

\author{Sabrina Maniscalco}
\affiliation{QTF Centre of Excellence, Turku Centre for Quantum Physics, Department of Physics and
Astronomy, University of Turku, FI-20014 Turun yliopisto, Finland}
\affiliation{7 QTF Centre of Excellence, Department of Applied Physics, Aalto University, FI-00076
Aalto, Finland}

\date{\today}

\begin{abstract}
The exploitation and characterization of memory effects arising from the interaction between system and environment is a key prerequisite for quantum reservoir engineering beyond the standard Markovian limit. In this paper we investigate a prototype of non-Markovian dynamics experimentally implementable with superconducting qubits. We rigorously quantify non-Markovianity highlighting the effects of the environmental temperature on the Markovian to non-Markovian crossover. We investigate how memory effects influence, and specifically suppress, the ability to perform work on the driven qubit. We show that the average work performed on the qubit can be used as a diagnostic tool to detect the presence or absence of memory effects.
\end{abstract}

\maketitle

\section{\label{sec:intro}Introduction}
Real quantum systems are always in contact with a surrounding environment, leading to the necessity of an open-system description \cite{breuer1,weiss}. 
Regardless of the fine microscopic details of the system-environment interaction, one can, to some extent, assess some general properties of the
reduced system dynamics  and, in particular, investigate the presence and role of memory effects \cite{breuer2,rivas1}.

In recent years, a substantial body of literature has focused on non-Markovian dynamics \cite{wolf,blp,rhp,luo,lorenzo1,bogna2}, also due
to relevant technological advances that have made it possible to observe shorter time scales at which memory effects do play an 
important role \cite{liu,li,cialdi,tang,chiuri,jin,orieux,bernardes, cialdi2017}. The range of approaches to defining and quantifying non-Markovian behaviour as well as  applications in quantum information and simulation
protocols \cite{antti1,biheng,laine2014}, foundational issues of quantum mechanics \cite{teiko} and even studies in condensed matter is incredibly vast \cite{lorenzo3,pinja1,sindona,cetina1,pinja2,massimo,lorenzo2}. Although no unique answer on the physical nature of non-Markovianity is currently available, its intrinsic multi-faceted aspect has surely sparked interest in a plethora of different fields.

In several physical models of open quantum systems, the Markovian (memory-less) or non-Markovian character of the dynamics is crucially connected to a physical parameter determining the relative time scales of the system-environment interaction. In the spirit of reservoir engineering one can, in certain physical implementations, manipulate such parameters driving and observing the Markovian to non-Markovian crossover. This has been done experimentally for simple models of qubit dynamics, mainly with optical setups \cite{liu,bernardes,cialdi2017}.

In this paper we consider a system consisting of a (driven) qubit coupled to a non-Markovian environment modelled as an additional qubit, which in turn dissipates to a Markovian thermal bath. For the sake of conveying the objectivity of qubits, in the rest of the paper we refer to the first qubit as the cold qubit (CQ) and to the latter qubit as the thermal qubit (TQ). The TQ plays here the role of the memory of the non-Markovian environment. Depending on the relative strength of the CQ-TQ coupling and the TQ-Markovian bath coupling, one observes the presence or absence of memory effects. One of our main goals is to establish a relationship between such a physical parameter and a recently introduced information-based quantifier of non-Markovianity, namely the volume of accessible states \cite{lorenzo1}. One would expect a linear increase of non-Markovianity when increasing the CQ-TQ coupling with respect to  the TQ-Markovian bath coupling. Interestingly, we find that this is not exactly the case as we observe a slightly non-monotonic behaviour.

We then turn our analysis to the thermodynamic properties of the open quantum system \cite{campisi,roncaglia,talkner}, and in particular we look at the connection between the average work performed on the CQ \cite{solinas,rebecca} and memory effects in the dynamics, as measured by the volume of accessible states. The reason for this specific choice of quantities lies in the fact that they are both conceptually easy to grasp and mathematically solid to investigate the possible interplay between quantum thermodynamics and non-Markovianity. Our main target is to understand if such an interplay exists, and whether it can be used to diagnose the presence of non-Markovian effects. In particular, we find that presence of memory effects suppresses the average of work performed on the CQ under resonant periodic driving. This finding, in addition to be used as a diagnostic tool for memory effects, can be of high importance in certain quantum thermodynamic tasks. If we want to increase (decrease) the ability to perform work on a qubit, then our results show that non-Markovian environments perform worse (better) than Markovian ones.

While our theoretical analysis is completely general and not specifically dependent on the physical context in which the model can be experimentally implemented, to illustrate our findings  we will use parameter values that are typical of the superconducting qubits scenario as we believe that this model can be soon in the grasp of experimentalists in this framework. We notice that a similar system has been studied by some of the authors in the classical regime where the  role of hidden variables in the observable (thermo)-dynamics  was studied \cite{mehl,massimo2}. Moreover, two-qubit system when each qubit interacts with its own bath is employed to study heat transport using numerically exact techniques in \cite{kato} and also quantum correlations in \cite{bellomo}.  

This manuscript is organised as follows. In Section~\ref{sec:model} the model we investigate is introduced. Section~\ref{sec:NM} is dedicated to a brief introduction of the tool we use to quantify non-Markovianity and its parameter dependence. In Sec.~\ref{sec:work} we study the work performed on a sub-part of the system in presence and absence of memory effects. Finally, we draw conclusions and discuss open perspectives in Sec.~\ref{sec:con}.

\section{\label{sec:model}The two-qubit model}

\begin{figure}[H]
\centering
 \includegraphics[width = 1 \columnwidth]{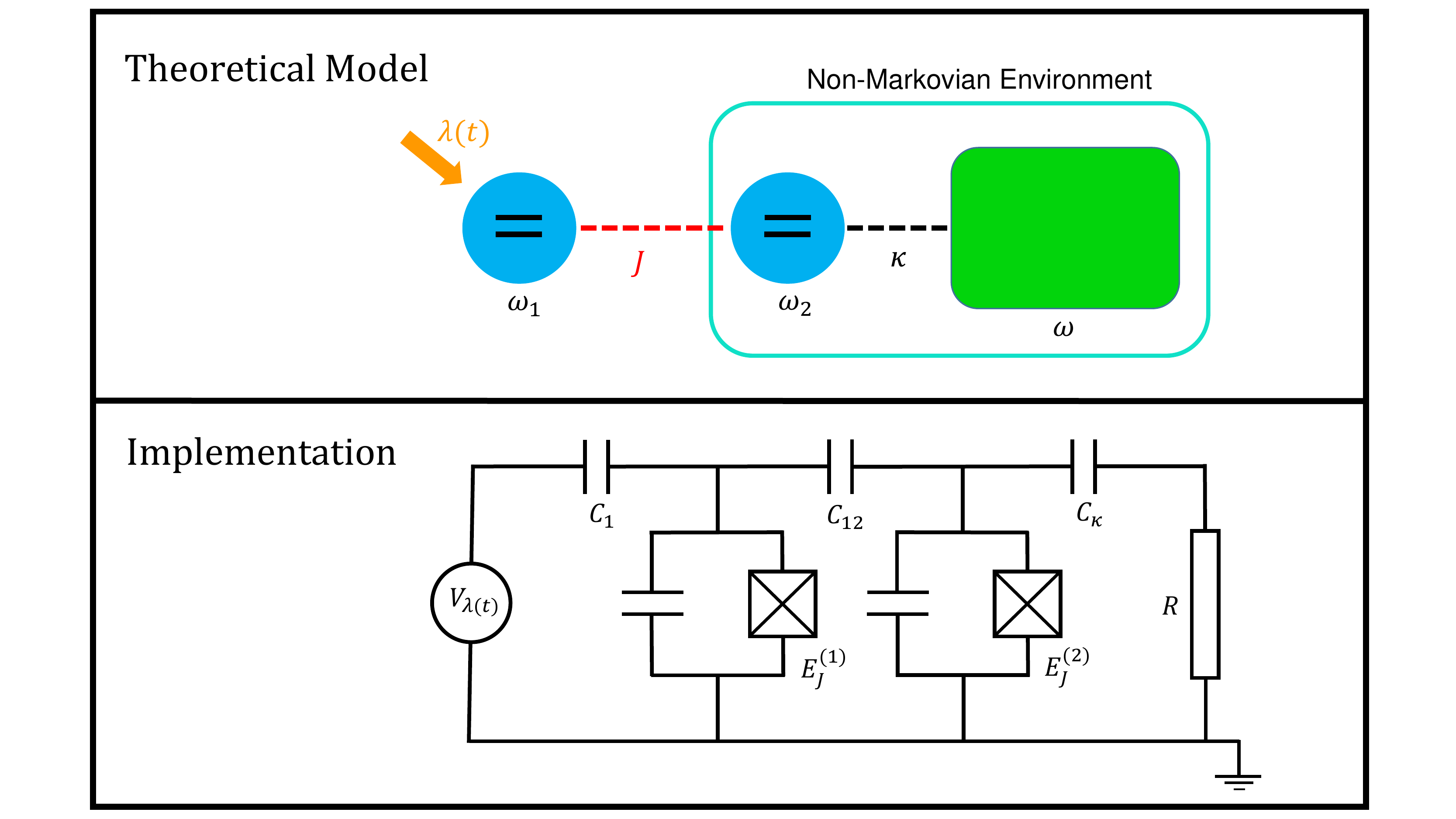}
\caption{Sketch of the theoretical model (upper panel) and a possible experimental implementation (lower panel) in circuits using transmon qubits. The CQ is coupled to a time-dependent voltage gate and capacitively coupled to a TQ. This, in turn, is connected to a resistor that acts as a dissipative environment.}
\label{scheme}
\end{figure}

As depicted in Fig.~\ref{scheme}, we consider two distinguishable interacting two-level systems, labeled 1 and 2, which respectively correspond to CQ and TQ. Whilst CQ is subjected to a weak periodic driving, TQ is weakly coupled to a Markovian non-degenerate bosonic bath at temperature $(\beta k_B)^{-1}$. This type of setting can be practically implemented using nano-devices operating in the quantum regime, such as flux qubits, Cooper pair boxes, and transmon qubits, and by
coupling one of them to a resistor~\cite{viisanen}. For the results presented in Secs. \ref{sec:NM} and \ref{sec:work}, we use parameters typical of such a superconducting setting, employing the engineered set-up to sweep through the range of parameters we study. Therefore, the typical energy scale of the qubit resonance frequency is about $\hbar\omega_{0}\approx 1 \enskip \textrm{K}\cdot k_{B}$. Beyond this experimental realization, the principal setup, as depicted in the upper panel of Fig.~\ref{scheme}, has been studied widely in the context of non-Markovianity. In modelling the non-Markovian features of the system-environment interaction we follow an approach conceptually similar to the pseudomode method, introduced in Ref. \cite{Garraway1996}, dividing the environment  into a part storing the memory (the TQ) and a memoryless Markovian part.  \\
The total Hamiltonian of the system reads (we assume $k_B=1$ and $\hbar= 1$ in the rest of the paper)
\begin{equation}
\begin{aligned}
H(t) &= \sum_{j=1}^{2}\frac{\omega_j}{2}\sigma^{(j)}_z + \sum_{\omega} \omega b^{\dagger}_{\omega}b_{\omega} + J\sigma^{(1)}_x\sigma^{(2)}_x \\
& + \lambda(t)\sigma^{(1)}_x+\sum_{\omega} \kappa g_\omega(\sigma^{(2)}_+b_\omega+\sigma^{(2)}_-b^{\dagger}_\omega),
\end{aligned}
\label{TotHam}
\end{equation}
in which $\omega_{j}$ is the frequency of the $j$-th qubit, $b_{\omega},b^{\dagger}_{\omega}$ are the bosonic annihilation and creation operators of the mode environment at frequency $\omega$, $J$ the qubit-qubit coupling constant, $\lambda(t)$ is a time-dependent driving protocol, $g_{\omega}$ is the spectral function of the environment and $\kappa$ is a dimensionless coefficient determining strength of the interaction between TQ and the bath. We consider a periodic driving field acting on CQ, specifically $\lambda(t)=\lambda_0 \sin[\omega_D t]$. In the lower panel of Fig. \ref{scheme} we depict a possible implementation using transmon qubits, whose dynamics can by described by \eqref{TotHam}, provided that a global $\pi$ rotation in the $x-y$ plane leading to $\sigma_x\to\sigma_y$, is performed. 
By considering the bipartite qubit system as an open system and limiting our attention to the weak driving regime, one can study the dynamics of the the joint 2-qubit density matrix $\rho(t)$ using the following Lindblad master equation
\begin{equation}
\dot{\rho}(t)=-i[H_S+H_D(t),\rho(t)]+\mathcal{D}[\rho(t)],
\label{MEPhen}
\end{equation}
in which we have relabelled $H_S=\sum_{j=1}^{2}\frac{\omega_j}{2}\sigma^{(j)}_z +J\sigma^{(1)}_x\sigma^{(2)}_x$ and $H_D(t)= \lambda(t)\sigma^{(1)}_x$, and $\mathcal{D}$ is the following Lindblad dissipator
\begin{eqnarray}
\mathcal{D}{\rho}(t)= &&\gamma ^{(\downarrow)} \Big(\sigma^{(2)}_{-}\rho(t)\sigma^{(2)}_{+}-\frac{1}{2}\{\sigma^{(2)}_{+} \sigma^{(2)}_{-},\rho(t)\}\Big) \nonumber
\\
&&+\gamma ^{(\uparrow)}  \Big(\sigma^{(2)}_{+}\rho(t)\sigma^{(2)}_{-}-\frac{1}{2}\{\sigma^{(2)}_{-} \sigma^{(2)}_{+},\rho(t)\}\Big),
\label{DissPhen}
\end{eqnarray}
in which $\gamma ^{(\downarrow)}=(\bar{\kappa}/2)[1+\coth(\omega_{2}\beta/2)]=\gamma^{(\uparrow)}e^{\beta\omega_{2}}$ are decay rates of the TQ, generally dependent upon the spectral density of the bath, which is assumed to be Ohmic with a cut-off frequency that is larger than all the relevant frequencies of the open system. Here we define $\bar{\kappa}=\kappa \omega_2$, which determines the TQ-bath interaction strength and is the zero-temperature decaying rate of the TQ.\\
The parameters identifying the relevant time scale are the coupling $J$ between the two qubits, the coupling $\bar{\kappa}$ between TQ and thermal bath (or more precisely the ratio between the former two), the qubit-qubit detuning $\Delta=\omega_1-\omega_2$, and the thermal time scale $\hbar\beta$.\\
A valid objection to using Eq.~\eqref{DissPhen} to describe the effect of the dissipative Markov bath could be raised by noticing that, if the qubits are strongly interacting, then a non-local dissipator
should be used to describe more realistically the dynamics of the combined two-qubit system. For this reason, we also derived a more complicated, non-local version of Eq.~\eqref{DissPhen} and employed it for comparison. However, for the parameter region we are interested in, no appreciable discrepancies between the two models were found and therefore, all the findings reported in the following are
obtained by using the local dissipator~\eqref{DissPhen} (see appendix). Additionally, we benchmarked our results with an exact, numerical method, the stochastic Liouville-von Neumann equation~\cite{SG02,rebecca}.

\section{Non-Markovian dynamics of the CQ}\label{sec:NM}
In this section we study how memory effects depend on the relevant time scales of our system. We are interested in the reduced dynamics of the CQ, which in this section is un-driven, i.e. $\lambda_{0}=0$. More specifically, having the reduced state of the CQ by taking the partial trace over the TQ, that is $\rho_{1}(t)=\textrm{tr}_{2}\rho(t)$, and considering the larger environment consisting of the TQ and the bath, non-Markovian features could arise in the reduced dynamics of the CQ.
To characterise the degree of non-Markovianity in the dynamics of qubit 1 we use the geometric approach introduced in \cite{lorenzo1}, which witnesses memory effects by monitoring the temporal evolution of the volume of accessible states during the dynamics. Other definitions are available in literature \cite{wolf,blp,rhp,luo,bogna2} and, depending on the system at hand, might be more or less suitable. The reason we chose
to utilise this quantifier lies in the relative computational simplicity for this particular system. It is worth mentioning that the volume of accessible states is a weak witness of non-Markovianity, in what it often fails to detect non-divisibility of the dynamical map. However, non-Markovianity detected by this approach always implies non-divisibility as well as back-flow of information \cite{DarSab}.\\

Consider a completely positive and trace-preserving (CPTP) map $\Lambda_t$ describing the evolution of the CQ, such that $\rho_1(t)=\Lambda_t[\rho_1(0)]$. One can find an affine transformation of the Bloch vector of the qubit, associated to this map, such that $\vec{r}(t)=A(t)\vec{r}(0)+\vec{T}(t)$. Here $\vec{r}(t)$ is the Block vector at a given time $t$, $A(t)$ is a transformation matrix  which rotates and possibly shrinks the Bloch vector, and $\vec{T}(t)$ is a translation of the Bloch vector. The volume of accessible states at a given time $dV(t)$ can be captured by evaluating the determinant of the transformation matrix, $\vert A(t)\vert$, \cite{lorenzo1}. 
\begin{equation}
dV(t)=\vert A(t)\vert dV(0),
\label{volume}
\end{equation}
in which $dV(0)$ is the initial differential of the volume of accessible states. It can be shown that for any Markovian (divisible) dynamical map $\vert A(t)\vert$ decreases monotonically in time. Therefore, a non-monotonic time-evolution of the volume of accessible states signal memory effects in the dynamics of an open system. This also allows one to define a quantifier
of the degree of non-Markovianity of a dynamical map as
\begin{equation}
\mathcal{N}=\frac{1}{V(0)}\int_{\dot{V}(t)>0}\dot{V}(t)dt,
\label{nonmark}
\end{equation}
in which the time-integration is performed over those time intervals accounting for the non-monotonicity of $V(t)$.
It is generally assumed for this widely studied generic setting we employ here, that the non-Markovianity increases monotonically with the ratio $J/\bar{\kappa}$. However, as we will show in the following, the behaviour of non-Markovianity for qubit 1 is slightly more sophisticated.

\begin{figure}
\centering
 \includegraphics[width = 0.8 \columnwidth]{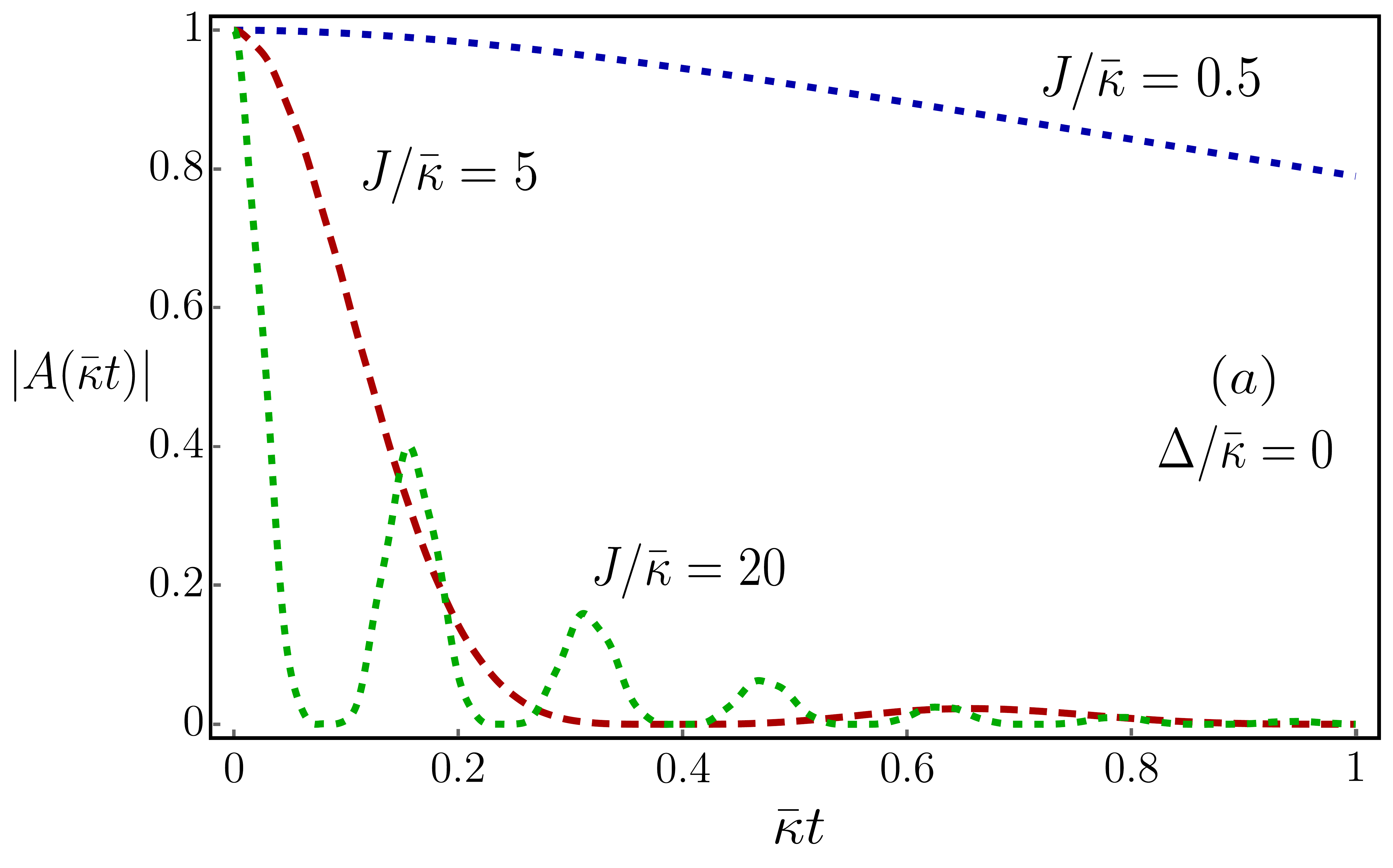}
\vspace{2 mm}
 \includegraphics[width = 0.8 \columnwidth]{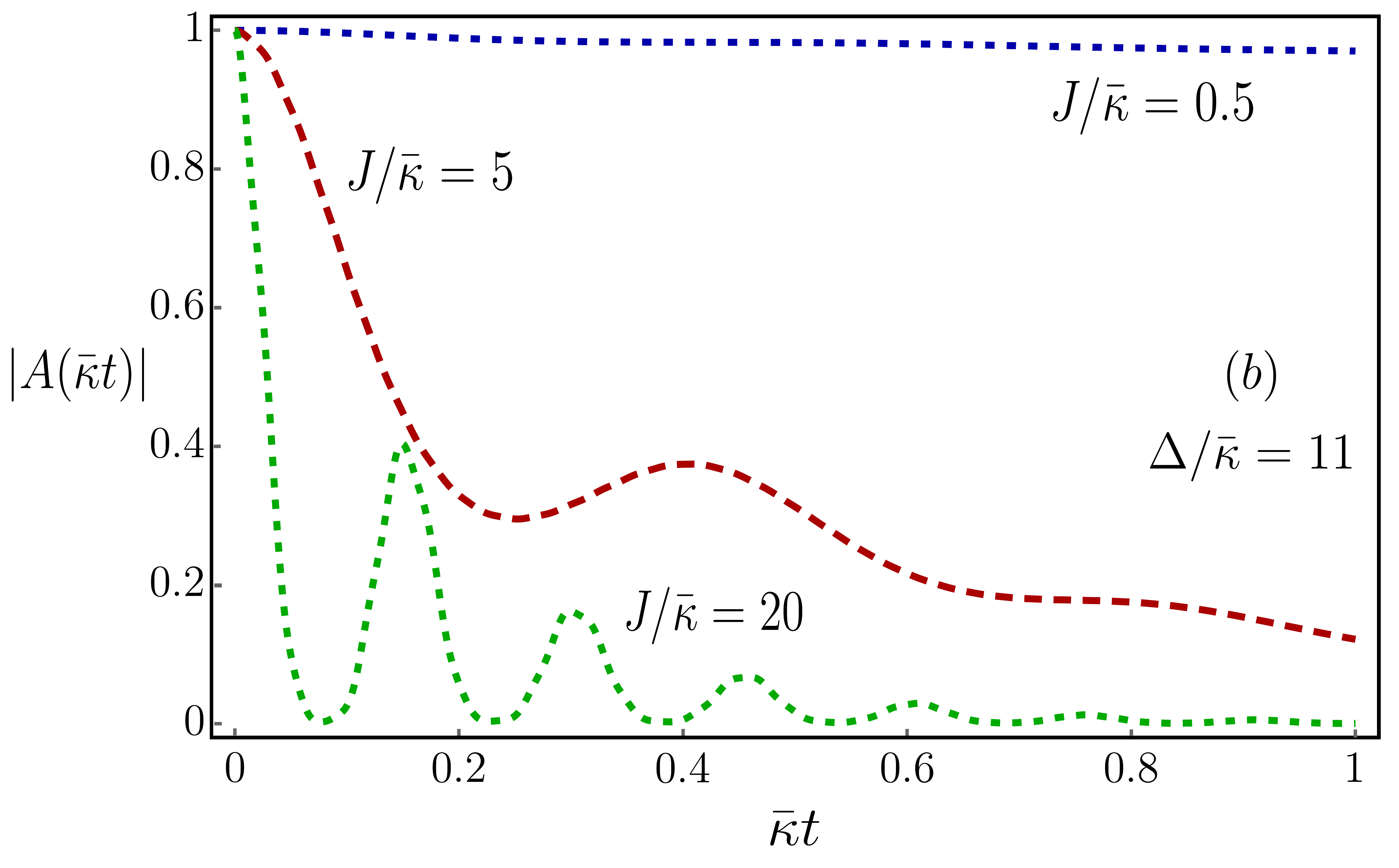}
\caption{Time-evolution of the volume of accessible states $|A(\bar{\kappa}t)|$ at low-temperature regime with $\omega_{2}\beta=0.35$, different values of $J/\bar{\kappa}$ and zero detuning ($\Delta/\bar{\kappa}=0$, a) and non-zero detuning ($\Delta/\bar{\kappa}=11$, b).}
\label{atplot}
\end{figure}

As the TQ is part of the environment, its initial state is thermalised, $\rho_{2}(0)=e^{-\beta\omega_{2}\sigma_{z}^{(2)}/2}/\mathcal{Z}_{2}$ with $\mathcal{Z}_{2}= \mathrm{tr}[e^{-\beta\omega_{2}\sigma_{z}^{(2)}/2}]$  . 
We investigate the two cases of resonant and non-resonant qubits. We start by inspecting the evolution of $|A(\bar{\kappa}t)|$ for several values of the tuning parameter $J/\bar{\kappa}$ at a fixed low temperature of $\omega_{2}\beta=0.35$, as displayed in Fig. \ref{atplot}. As expected, a transition from Markovian to non-Markovian dynamics occurs at a certain threshold value $(J/\bar{\kappa})_\textrm{th}$ at which the influence of the TQ (TLS), carrying the memory, becomes dominant over the Markovian thermal bath. So we define $(J/\bar{\kappa})_\textrm{th}$ as the smallest value of the $J/ \bar{\kappa}$ at which the volume of accessible states,
$|A(t)|$, shows temporary increase, i.e., starts to display non-monotonic behaviour. Depending on whether the qubits are resonant or not, $(J/\bar{\kappa})_\textrm{th}$ can be larger or smaller. This can be qualitatively understood using the following argument. At low temperatures, when $\Delta/\bar{\kappa}=0$, the thermal bath is effectively in resonance with both qubits. Therefore, being this made of a small density of bosonic modes thermally excited around $\omega_{1}$, it tends to overrule the effects of the TQ. This is valid up to a critical value of $J/\bar{\kappa}$ at which, the qubit-qubit interaction is strong
enough for memory effects to kick in. When the two qubits are non-resonant, the effective action of the thermal bath on the CQ becomes strongly off-resonant for lower values of $J/\bar{\kappa}$. The Markov bath becomes energetically transparent to the CQ, which at this point, is effectively interacting with the TLS only. 

\begin{figure}
\centering
 \includegraphics[width = 0.95 \columnwidth]{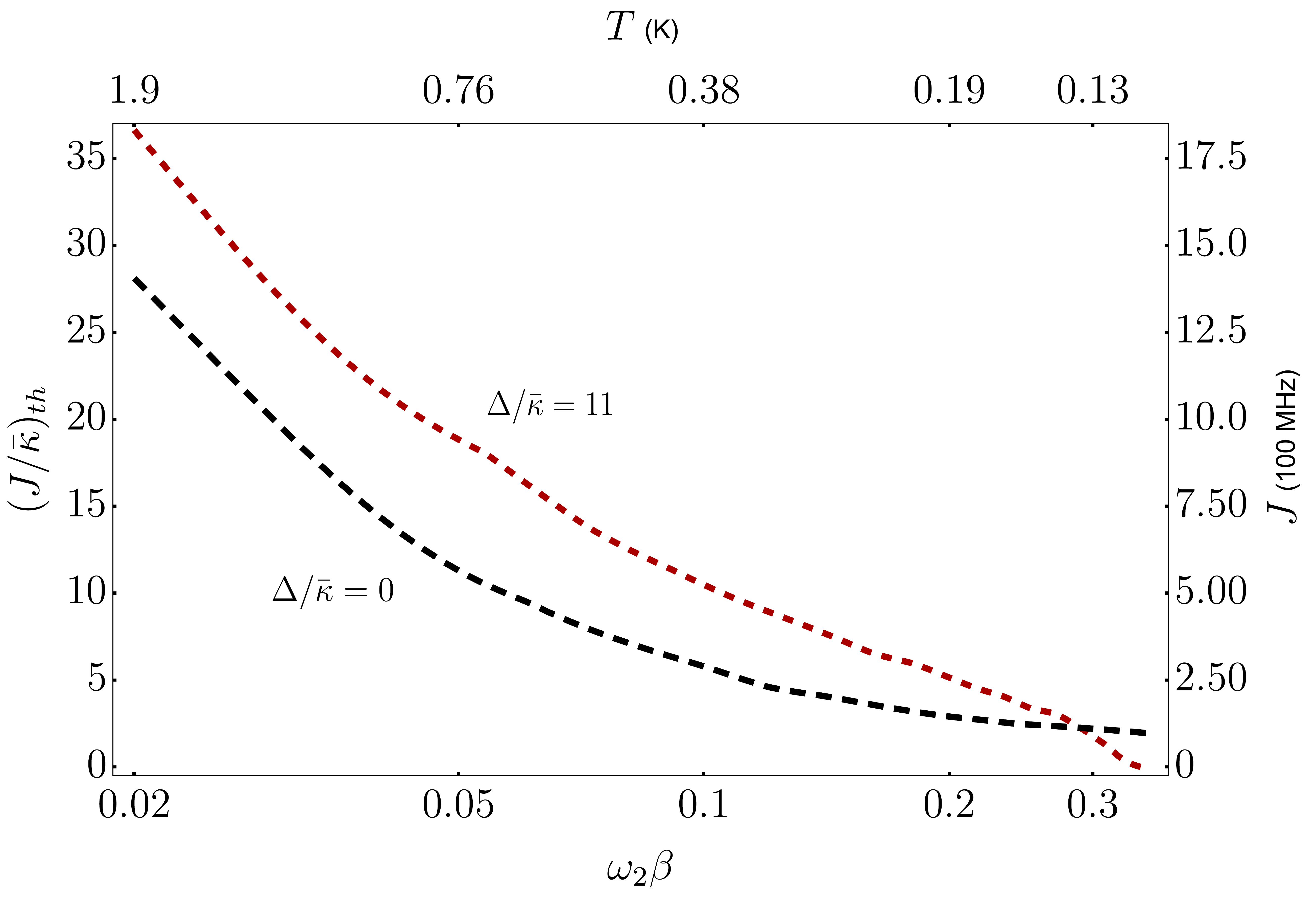}
\caption{Threshold value of $J$ for Markovian to Non-Markovian crossover in $Y$ axes versus temperature of the Markovian bath in $X$ axes in $log$ scale for resonant and non-resonant qubits. Left $Y$ and lower $X$ axes show respectively the threshold $J/\bar{\kappa}$ and rescaled temperature of the bath, $\omega_2\beta$. Right $Y$ and upper $X$ axes determine threshold value of $J$ in the unit of $100$ MHz and temperature of the Markovian bath in Kelvin, respectively, provided that $\omega_2=5$ GHz and $\bar{\kappa}=50$ MHz. }
\label{jthplot}
\end{figure}

\begin{figure}
\centering
 \includegraphics[width = 0.8 \columnwidth]{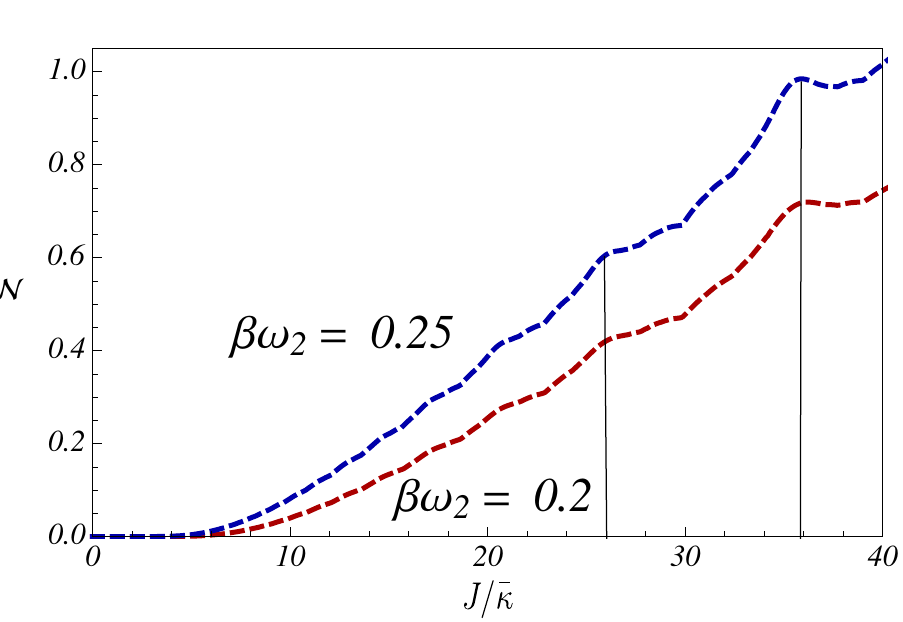}
\caption{Non-Markovianity measure $\mathcal{N}$ for resonant qubits as a function of $J/\bar{\kappa}$ for two different rescaled temperature $\omega_{2}\beta$.}
\label{nmplot}
\end{figure}

This argument works well in the low-temperature limit but it fails in the high-temperature limit, as shown by Fig. \ref{jthplot}, where the threshold values $(J/\bar{\kappa})_\textrm{th}$ is plotted versus $\omega_2\beta$ for both the $\Delta/\bar{\kappa}=0$ and $\Delta/\bar{\kappa}=11$ cases.
While in the low-temperature limit, that is $\omega_2\beta\to\infty$, the threshold value $(J/\bar{\kappa})_\textrm{th}$ tends asymptotically to 1 for resonant qubits and to 0 for non-resonant qubits, if the rescaled temperature $\omega_2\beta$ is chosen above a certain value the situation is reversed and the threshold coupling for non-resonant qubits is larger than the one for resonant ones. At higher temperatures the density of thermally excited modes increases. With the qubit-qubit detuning being relatively small, the frequency of the CQ will be resonant with a bath mode that is sufficiently thermally occupied and will couple to it via interaction with the TQ. Therefore, since the TQ is off-resonance with the CQ, it will take a larger interaction strength $J$ to induce memory effects on the reduced dynamics of the latter. One can easily fit the $\Delta/\bar{\kappa}=0$ curve to a power-law decay 
\begin{equation}
\left(\frac{J}{\bar{\kappa}}\right)_\textrm{th}-1\approx \frac{A}{(\omega_2 \beta)^{B}}
\label{powerlaw}
\end{equation}
in which $A$ and $B$ can be numerically determined. Now we give some insight into the order of magnitude of the threshold $J$ regarding some typical experimental values. Suppose that two qubits are resonant at the frequency $\omega_1=\omega_2=5$ GHz and the zero-temperature decaying rate of TQ is $\bar{\kappa}=50$ MHz. According to Fig. \ref{jthplot} the threshold value of the $J$ for low temperatures asymptotically approaches $50$ MHz, which is a typical value for qubit-qubit coupling in experimental setups using transmon qubits \cite{ronzani}. It is also worth mentioning that the value of $\bar{\kappa}$ can be determined experimentally, provided that we know the temperature of the bath, by detecting the relaxation time of the qubit via a dispersive measurement of the qubit state following a driving pulse \cite{masuyama}.\\
We conclude this section by inspecting the behaviour of $\mathcal{N}$ vs $J/\bar{\kappa}$ at fixed temperature, as displayed in Fig. \ref{nmplot} for $\beta\omega_{2}=0.2,0.25$. First, we notice, in agreement with Fig. \ref{atplot}, the presence of a threshold value $(J/\bar{\kappa})_{th}$. Second, the memory effects become stronger at lower temperatures, a trend consistent with the low-temperature
argument introduced above. As anticipated, we find that non-Markovianity is not monotonically increasing with $J/\bar{\kappa}$: local extrema can be observed whose position in $J/\bar{\kappa}$ appears to be independent of the chosen (low) temperature of the bath.\\ 
\section{Average work}\label{sec:work}
\begin{figure}
\centering
 \includegraphics[width = 0.8 \columnwidth]{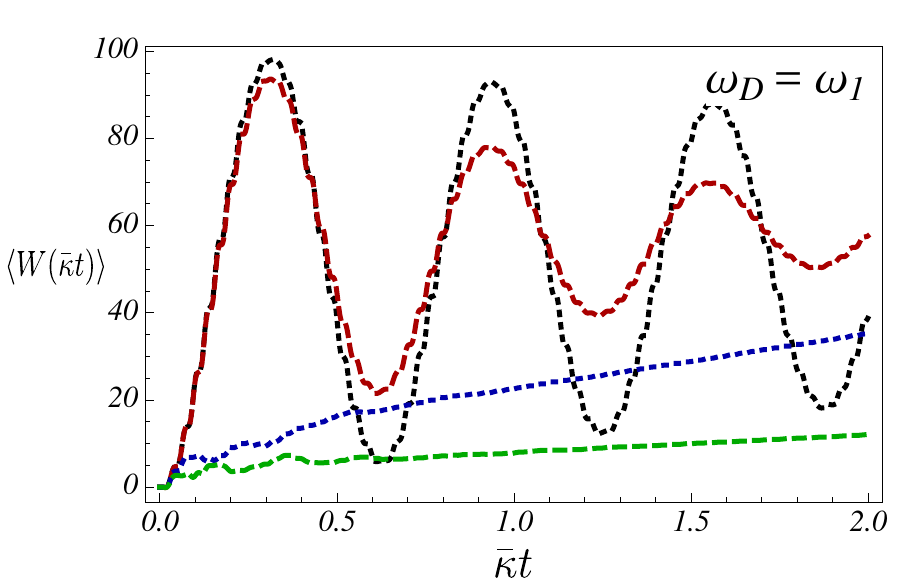}
\vspace{2 mm}
 \includegraphics[width = 0.8 \columnwidth]{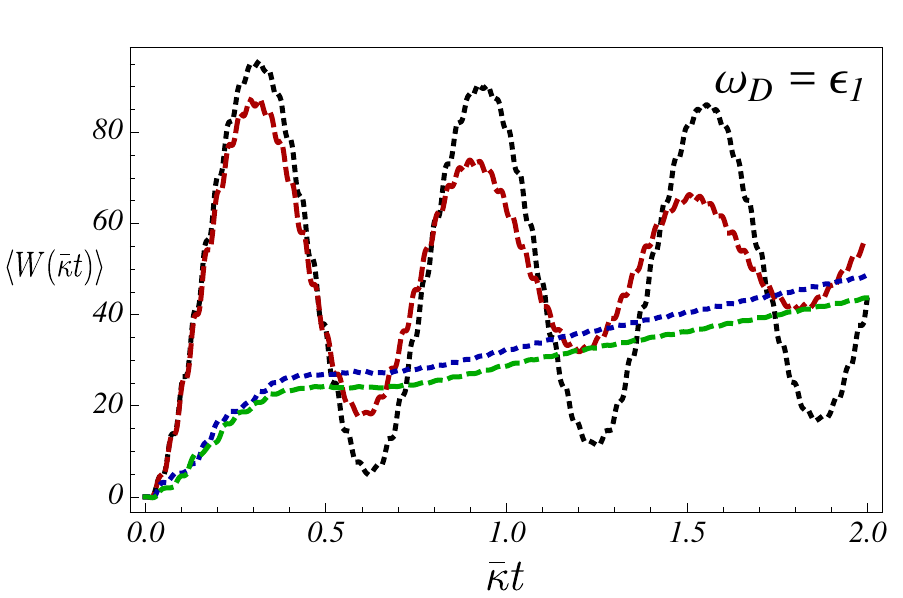}
\caption{Time-evolution of the average work performed on the CQ at temperature $\omega_{2}\beta=0.2$, $\Delta/\bar{\kappa}=0$ and for the following values of $J/\bar{\kappa} =1$ (dotted black),2 (dashed red), 20 (dotted blue), 30 (dashed green). In the upper panel the frequency of the driving field is set to $\omega_{D}=\epsilon_{1}$ while in the lower panel  $\omega_{D}=\omega_{1}$.}
\label{wplot}
\end{figure}

In this section, we include in the total system the weak driving force $H_{D}(t)$ acting on the CQ and we investigate how memory-effects arising in the 
un-driven dynamics can affect the average work performed. Driving changes in general the non-Markovianity of an open system, it can be generated, enhanced, partly or fully suppressed~\cite{RS16,Basilewitsch17,Sampaio17}. While we studied the non-Markovianity both for the driven and un-driven qubit, for the effect discussed, the non-Markovianity of the undriven qubit turns out to be the most relevant. Also, we restrict both the coupling strength $J$ and the driving amplitude $\lambda_0$ to small values, consistently with the approximations done to derive our master equation.\\
The problem of defining work in the quantum realm has attracted a great deal of attention recently~\cite{campisi,roncaglia,talkner,solinas}. Due to the foundational problem posed by quantum measurement
and the impossibility of defining work as an observable, several approaches have been considered, mostly limited to the case of unitary dynamics. A
significant and well defined quantity is the average of the work performed, introduced in \cite{solinas} and applicable to the case of open systems too.
In this setting, the average work is defined via the power operator
\begin{equation}
P(t)\equiv\frac{\partial H}{\partial  t}(t),
\label{power}
\end{equation}
which is linked to the first momentum of work via the following relation
\begin{equation}
\langle W(t)\rangle=\int_0^t d \tau \langle P(\tau) \rangle.
\end{equation}
The interplay between non-Markovianity and weak driving can lead to interesting results.
We assume the CQ is driven by a resonant periodic driving protocol $\lambda(t)=\lambda_0 \sin[\omega_D t]$ that changes its free Hamiltonian in time. \\
Since the two qubits are at all times interacting one may wonder what is the correct frequency $\omega_{D}$ at which the CQ should be driven.  Generally speaking,
due to the qubit-qubit interaction, the natural oscillation frequencies of the total system are dressed. On the other hand, since we are working
in a weak coupling regime, this frequency dressing should not be too drastic and one should still be able to consider the two qubits as separated entities, at least to some extent.
Therefore, an interesting question naturally arises on how the work performed on the CQ changes depending not only on the presence of memory effects, but also
on the specific driving frequency. For all the above reasons, we choose two different driving frequencies, the bare CQ transition frequency $\omega_{1}$ and the lower
non-degenerate transition frequencies of the joint qubit-qubit system $\epsilon_{1}=\sqrt{\omega_{2}^{2}+J^{2}}-J$.\\
In Fig. \ref{wplot} we display the time-evolution of the average work performed on the single qubit for these two driving frequencies. In both cases
the first striking feature is a suppression (damping of the oscillations) of the $\langle W(\bar{\kappa} t)\rangle$ when the CQ dynamics transitions from Markovian to 
non-Markovian. Given the sinusoidal driving we employ, one would expect an oscillatory behaviour of the average work, damped of course because of the bath attached. In fact, for longer propagation times (not shown), the oscillations vanish due to damping. Which is exactly the behaviour we observe in case of Markovian dynamics (red and black lines). But for non-Markovian case (blue and green lines), the average work does not evolve according to this expectation. Memory effects in the single qubit dynamics act as a friction that opposes to the ability to coherently drive the Bloch vector of the CQ. 
This effect appears stronger in the case of bare-frequency driving (upper panel). Although for stronger qubit-qubit interaction (blue and green line) one might argue
that this is due to off-resonant driving, the lower panel, where $\omega_{D}$ is set equal to $\epsilon_{1}$ and therefore $J-$dependent, seems to indicate otherwise.
At longer times, for both driving protocols, the average work increases linearly in time, suggesting the onset of a time-dependent steady state. Obviously, for
$\omega_{D}=\omega_{1}$ the stronger $J$ the more off-resonant the driving is, resulting in an average work barely increasing above zero. The suppression of work effect does not depend on the amount of non-Makovianity and can therefore be understood a qualifier for non-Markovianity, which is experimentally accessible.  \\
\section{Conclusions}\label{sec:con}
In this manuscript we have addressed the interplay between memory effects and the work performed on part of a larger quantum system. The system considered here consists of two-interacting qubits; one of them is coupled to a thermal bath and the other one is driven coherently. We have used a quantifier of memory effects based on the volume of accessible states and found a threshold value for the relevant couplings at which single-qubit non-Markovian dynamics sets in. Furthermore, we have shown that memory effects induce a significant suppression of the work performed on the driven qubit under resonant periodic driving. Besides shedding light on the interplay between non-Markovianity and out-of-equilibrium dynamics, this findings suggest a novel diagnostic tool for testing the degree of non-Markovianity in some experimentally interesting scenarios. 

\section*{Acknowledgements}
S. H.-R., S.M., R. S. and M. B. acknowledge financial support from the Horizon 2020 EU collaborative projects QuProCS (Grant Agreement No. 641277),  the Academy of Finland (Project no. 287750), and the Magnus Ehrnrooth Foundation. This work was performed as  part of the Academy of Finland Centre of Excellence program (projects 312057 and 312058).
\section*{Appendix: Comparison between a non-local and a local master equation}
\setcounter{equation}{0}
\setcounter{figure}{0}
\renewcommand{\thefigure}{A\arabic{figure}}
\renewcommand{\theequation}{A\arabic{equation}}
We recast the total Hamiltonian of the CQ-TQ system and the bath as $H(t)=H_S+H_B+H_I+H_D(t)$, where $H_S$ is the inner Hamiltonian of the CQ-TQ system, $H_B$ is the Hamiltonian of the bath, $H_I$ is the interaction Hamiltonian , and $H_D(t)$ is the driving Hamiltonian. A microscopic derivation \citep{Petroccione} that accounts for Born-Markov approximation and weak driving leads to the following equation
\begin{eqnarray}\label{MEgeneral}
\frac{d}{dt}\tilde{\rho}(t)=&&-i[\tilde{H}_D(t),\tilde{\rho}(t)]
\\
&&-\int^{\infty}_0 ds \mathrm{tr}_B\Big\{[\tilde{H}_I(t),[\tilde{H}_I(t-s),\tilde{\rho}(t)\otimes \rho_B]]\Big\},\nonumber
\end{eqnarray}
where $\rho_B$ is the stationary state of the bath and $\tilde{\rho}(t)$ denotes density operator of the open system in the interaction picture. Note that the contribution of the driving Hamiltonian is neglected in the second term of \eqref{MEgeneral}. This is justifiable because of the Born-Markov approximation and that $H_D(t)$ is a local operator. The commutators can be expanded by recasting the TQ-bath interaction as
\begin{equation}
H_I=\sigma_x\otimes B_x+\sigma_y\otimes B_y,
\end{equation}  
where $\sigma_{x,y}$ are Pauli matrices of the TQ and  $B_x=\frac{\kappa\Sigma_{\omega}g_{\omega}(b_{\omega}+b^{\dagger}_{\omega})}{2}$, $B_y=\frac{i\kappa\Sigma_{\omega}g_{\omega}(b_{\omega}-b^{\dagger}_{\omega})}{2}$. The exact eigenvalues and eigenstates of $H_S$ can be easily computed
 \begin{eqnarray}
\ket{E_4}&&=\alpha \ket{e}\ket{e}+\xi\ket{g}\ket{g}, \quad E_4 =\frac{1}{2}\sqrt{4J^2+\Omega^2}, 
\\
\ket{E_3}&&=\eta \ket{e}\ket{g}-\delta \ket{g}\ket{e}, \quad E_3=\frac{1}{2}\sqrt{4J^2+\Delta^2},
\\
\ket{E_2}&&=\eta\ket{g}\ket{e}+\delta \ket{e}\ket{g}, \quad E_2=-E_3,
\\
\ket{E_1}&&=\alpha\ket{g}\ket{g}-\xi \ket{e}\ket{e}, \quad E_1=-E_4,
\end{eqnarray}
in which we have defined $\Omega=\omega_1+\omega_2$ and $\Delta=\omega_1-\omega_2$ and the coefficients $\alpha, \delta, \eta, \xi$  depend upon the parameters of the model as follows 
\begin{eqnarray}
\alpha=\frac{\Omega+\sqrt{4J^2+\Omega^2}}{\sqrt{(\Omega+\sqrt{4J^2+\Omega^2})^2+4J^2}}, 
\\
\xi=\frac{2J}{\sqrt{(\Omega+\sqrt{4J^2+\Omega^2})^2+4J^2}},
\\
\eta=\frac{\Delta+\sqrt{4J^2+\Delta^2}}{\sqrt{(\Delta+\sqrt{4J^2+\Delta^2})^2+4J^2}}, 
\\
\delta=\frac{-2J}{\sqrt{(\Delta+\sqrt{4J^2+\Delta^2})^2+4J^2}}.
\end{eqnarray}
The above eigenstates can be used to define the following global, CQ-TQ Lindblad operators
\begin{eqnarray}
L_x(\epsilon_1)&&=(\alpha \eta-\xi \delta)(\ket{E_3}\bra{E_4}+\ket{E_1}\bra{E_2}),\label{LindbladOp1}
\\
L_y(\epsilon_1)&&=i(\alpha \eta+\xi \delta)(\ket{E_3}\bra{E_4}+\ket{E_1}\bra{E_2}),\label{LindbladOp2}
\\
L_x(\epsilon_2)&&=(\alpha \delta+\xi \eta)(\ket{E_2}\bra{E_4}-\ket{E_1}\bra{E_3}),\label{LindbladOp3}
\\
L_y(\epsilon_2)&&=i(\alpha \delta-\xi \eta)(\ket{E_2}\bra{E_4}-\ket{E_1}\bra{E_3}),\label{LindbladOp4}
\end{eqnarray}
in which $\epsilon_1$ and $\epsilon_2$ are the qubit-TLS non-degenerate energy gaps 
{\small \begin{eqnarray}
&&\epsilon_1=\frac{1}{2}(\sqrt{4J^2+\Omega^2}-\sqrt{4J^2+\Delta^2}),
\\
&&\epsilon_2=\frac{1}{2}(\sqrt{4J^2+\Omega^2}+\sqrt{4J^2+\Delta^2}).
\end{eqnarray}}
After some standard steps one arrives to the following non-local master equation
\begin{eqnarray}
\frac{d}{dt}\tilde{\rho}(t)&&=
\\
&&-i[\tilde{H_D}(t),\tilde{\rho}(t)]+\Big\{\sum_{\nu ,\mu}\sum_{i,j}\e^{i t (\nu-\mu)}\Gamma_{ji}(\nu)\nonumber
\\
&&
\Big(L_i(\nu)\tilde{\rho}(t)L_j(\mu)^{\dagger}-L_j(\mu)^{\dagger}L_i(\nu)\tilde{\rho}(t)\Big)+\mathrm{h.c.}\Big\},\nonumber
\label{MEbeyondRWA}
\end{eqnarray}
with $\Gamma_{ji}(\nu)=\int_0^{\infty}ds \e^{i\nu s}\mathrm{tr}_B[\tilde{B}_j^{\dagger}(t)\tilde{B}_i(t-s)]$. \\
When performing the secular approximation that leads to a Lindblad master equation, one drops terms oscillating much faster than the shortest relaxation time of the open system. The corresponding time scale is expressed by the decay rates of Eq.~\eqref{MEbeyondRWA}, given by $\gamma(\nu)=\Gamma_{ji}(\nu)+\Gamma_{ij}^{*}(\nu)$ . This means that a Markovian master equation in Lindblad form can be derived only when $\min\{2\epsilon_1,2\epsilon_2, \epsilon_2-\epsilon_1\}\gg \max\{\gamma(\epsilon_1),\gamma(\epsilon_2)\}$. Provided that this requirement is fulfilled, after neglecting all oscillating terms and switching back to the Schr\text{\"o}dinger picture one finds
\begin{eqnarray}\label{LindbladME}
\dot{\rho}(t)&&=-i[H_S+H_D(t),\rho(t)] 
\\
&&+\gamma^{(\downarrow)}({\epsilon_1}) \Big(L_1\rho(t)L^{\dagger}_1-\frac{1}{2}\{L^{\dagger}_1 L_1,\rho(t)\}\Big) \nonumber
\\ 
&&+\gamma^{(\uparrow)}(\epsilon_1) \Big(L^{\dagger}_1\rho(t)L_1-\frac{1}{2}\{L_1 L^{\dagger}_1,\rho(t)\}\Big) \nonumber
\\
&&+\gamma^{(\downarrow)}({\epsilon_2}) \Big(L_2\rho(t)L^{\dagger}_2-\frac{1}{2}\{L^{\dagger}_2 L_2,\rho(t)\}\Big) \nonumber
\\
&&+\gamma^{(\uparrow)}(\epsilon_2) \Big(L^{\dagger}_2\rho(t)L_2-\frac{1}{2}\{L_2 L^{\dagger}_2,\rho(t)\}\Big), \nonumber
\end{eqnarray}
where we have defined $L_{1(2)}=1/2\big(L_x(\epsilon_{1(2)})-iL_y(\epsilon_{1(2)})\big)$ and considered a notation in which $\gamma^{(\downarrow)}(\epsilon_i)=\gamma( \epsilon_i )$ and $\gamma^{(\uparrow)}(\epsilon_i)=\gamma(- \epsilon_i)$.\\

The reliability of Eq.~\eqref{LindbladME} is dependent upon the validity of the secular approximation. In particular, this means that for resonant qubits ($\omega_1=\omega_2$) the condition $2J\gg\max\{\gamma^{(\downarrow)}(\epsilon_1),\gamma^{(\downarrow)}(\epsilon_2)\}$ must be fulfilled. This requirement sets a lower limit on the value of $J$. Thus, when the qubits are resonant, Eq. \eqref{LindbladME} is valid for a stronger CQ-TQ interaction. On the other hand, when qubits are resonant and $J$ is small, a valid master equation must contain those terms that are neglected in the secular approximation and therefore cannot be in the Lindblad form with respect to the operators (\ref{LindbladOp1}-\ref{LindbladOp4}).\\

On the other hand, the local master equation we use in the manuscript reads
\begin{eqnarray}\label{MEPhenA}
\dot{\rho}(t)=&&-i[H_S+H_D(t),\rho(t)]
\\
&&+\gamma^{(\downarrow)}(\omega_2) \Big(\sigma^{(2)}_{-}\rho(t)\sigma^{(2)}_{+}-\frac{1}{2}\{\sigma^{(2)}_{+} \sigma^{(2)}_{-},\rho(t)\}\Big) \nonumber
\\
&&+\gamma^{(\uparrow)}(\omega_2)  \Big(\sigma^{(2)}_{+}\rho(t)\sigma^{(2)}_{-}-\frac{1}{2}\{\sigma^{(2)}_{-} \sigma^{(2)}_{+},\rho(t)\}\Big), \nonumber
\end{eqnarray}
whose dissipating part is local Lindblad in the TQ. To understand the range of validity of this master equation, we compare it to  \eqref{MEbeyondRWA}. Let us expand the local Lindblad operators, $\sigma_+$ and $\sigma_-$, with respect to the non-local operators (\ref{LindbladOp1}-\ref{LindbladOp4}), such that 
\begin{equation}
\sigma_-=\frac{1}{2}\sum_{j=1,2}\Big(L_x(\epsilon_j)-iL_y(\epsilon_j)+L_x(\epsilon_j)^{\dagger}-iL_y(\epsilon_j)^{\dagger}\Big).
\end{equation}
Now we substitute this expression and its conjugate transpose in the \eqref{MEPhenA} and compare the resulting equation with the ME \eqref{MEbeyondRWA} in the Schr\text{\"o}dinger picture. Reminding the definitions of $\Gamma_{ji}(\nu)$ and $\gamma(\nu)$, it is straightforward to check that two master equqations coincide when  $\gamma^{(\downarrow)}(\epsilon_1)=\gamma^{(\downarrow)}(\epsilon_2)=\gamma^{(\downarrow)}(\omega_2)$ and $\gamma^{(\uparrow)}(\epsilon_1)=\gamma^{(\uparrow)}(\epsilon_2)=\gamma^{(\uparrow)}(\omega_2)$. Therefore, the local master equation can be justified microscopically when the decay rates act collectively. We stress that while Eq.~\eqref{MEPhenA} is in Lindblad form with respect to the local operators $\sigma_{\pm}$ of the TQ, it contains all the rotating terms with respect to the non-local operators (\ref{LindbladOp1}-\ref{LindbladOp4}). It is noteworthy to mention that a similar statement is presented in \citep{ScalaIOP} for a different but somewhat related physical model, that is the dissipative Jaynes-Cummings Model.
\begin{figure}\label{error}
\includegraphics[width=0.8\linewidth]{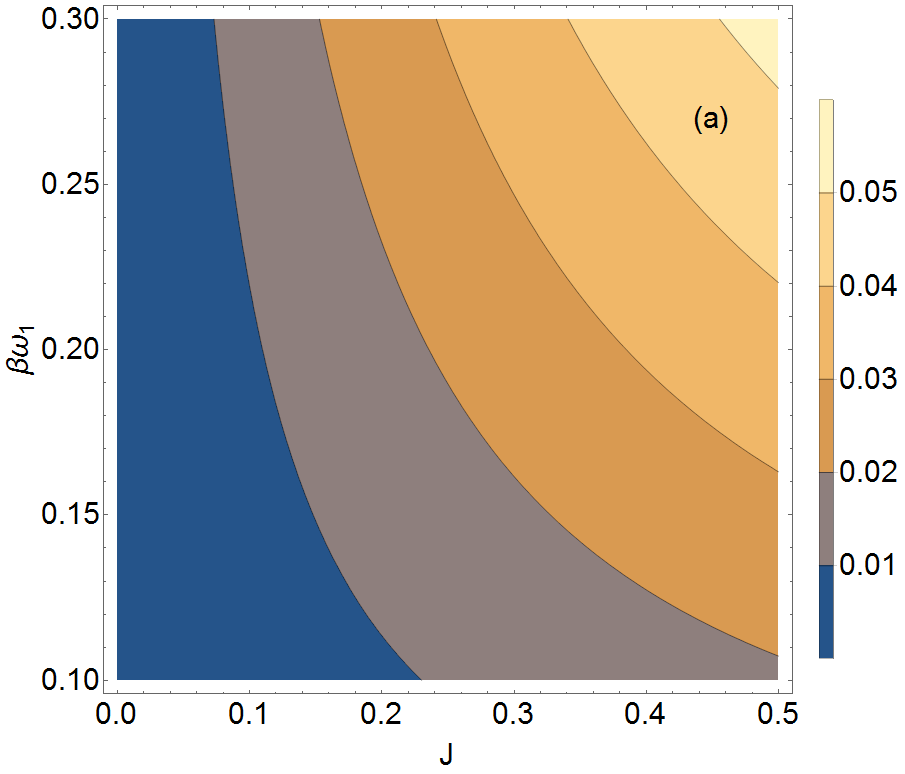}
\includegraphics[width=0.8\linewidth]{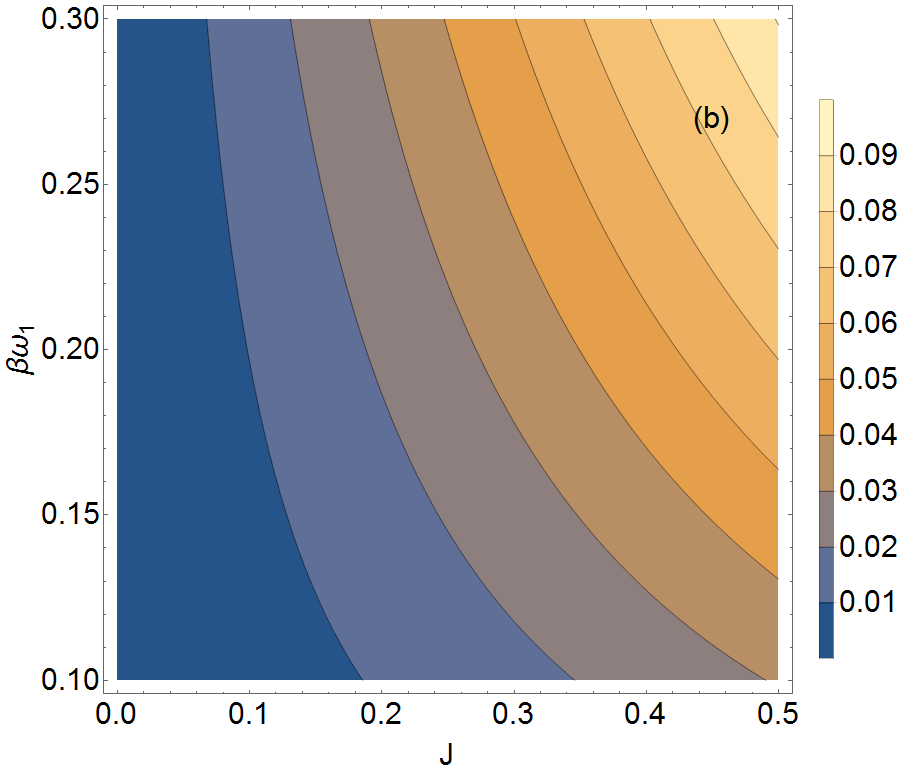}
\caption{Relative error of replacing decay rates of the two-qubit system, $\gamma^{(\downarrow)}(\epsilon_i)$, by the decay rates of the TQ, $\gamma^{(\downarrow)}(\omega_2)$, when qubits are in resonance. Respectively $\vert \frac{\gamma^{(\downarrow)}(\epsilon_1)-\gamma^{(\downarrow)}(\omega_2)}{\gamma^{(\downarrow)}(\omega_2)}\vert$ and $\vert \frac{\gamma^{(\downarrow)}(\epsilon_2)-\gamma^{(\downarrow)}(\omega_2)}{\gamma^{(\downarrow)}(\omega_2)}\vert$ are plotted in (a) and (b) as a function of $J$ and $\beta \omega_1$ with $\omega_1=1$.} 
\end{figure}
\\
Accordingly, one may approximate the Eq.~\eqref{MEbeyondRWA} by the Eq.~\eqref{MEPhenA} when the differences between the decay rates are small enough. Considering Ohmic spectral density of the bath, decay rates can be defined by $\gamma^{(\downarrow)}(\omega)=(\kappa/2) \omega\big[1+\coth(\omega \beta/2)]=\gamma^{(\uparrow)}(\omega)\e^{\omega \beta}$. For resonant qubits, the difference between $\epsilon_1$ and $\epsilon_2$ is of the order of $2J$. The relative error of replacing $\gamma^{(\downarrow)}(\epsilon_1)$ and $\gamma^{(\downarrow)}(\epsilon_1)$ by $\gamma^{(\downarrow)}(\omega_2)$ is depicted in the Fig. A1 as a function of $J$ and $\beta \omega_1$. This figure clearly shows that the deviation of two decay rates from $\gamma^{(\downarrow)}(\omega_2)$ is small in the region of small to moderate values of $J$. This is also relevant in the case of high-temperature bath.


\end{document}